\documentclass[%
 reprint,
 amsmath,amssymb,
 aps,
]{revtex4-2}

\usepackage{graphicx}
\usepackage{dcolumn}
\usepackage{bm}
\usepackage{amsmath}
\usepackage{amssymb}
\usepackage{txfonts}
\usepackage{mathrsfs}
\usepackage{xcolor}
\usepackage[mathlines]{lineno}
\usepackage{booktabs}

\def\b0{{\mathbf 0}}

\def\b0{{\mathbf 0}}

\def\beq{\begin{equation}}
\def\eeq{\end{equation}}
\def\beqn{\begin{eqnarray}}
\def\eeqn{\end{eqnarray}}

\begin{document}

\title{Thermal Casimir effect in the spin-orbit coupled Bose gas}
\author{Marek Napiórkowski  }
\email{marek.napiorkowski@fuw.edu.pl}
\affiliation{Institute  of Theoretical Physics, Faculty of Physics, University of Warsaw, Pasteura 5, 02-093 Warsaw, Poland}
\author{Pawel Jakubczyk } 
\email{pawel.jakubczyk@fuw.edu.pl}
\affiliation{Institute of Theoretical Physics, Faculty of Physics, University of Warsaw, Pasteura 5, 02-093 Warsaw, Poland}

\date{\today}

\begin{abstract} 
We study the thermal Casimir effect in ideal Bose gases with spin-orbit (S-O) coupling of Rashba type below the critical temperature for Bose-Einstein condensation. In contrast to the standard situation involving no S-O coupling, the system exhibits long-ranged Casimir forces both in two and three dimensions ($d=2$ and $d=3$). We identify the relevant scaling variable involving the ratio $D/\nu$ of the separation between the confining walls $D$ and the S-O coupling magnitude $\nu$. We derive and discuss the corresponding scaling functions for the Casimir energy. In all the considered cases the resulting Casimir force is attractive and the S-O coupling $\nu$ has impact on its magnitude. In $d=3$ the exponent governing the decay of the Casimir force becomes modified by the presence of the S-O coupling, and its value depends on the orientation of the confining walls relative to the plane defined by the Rashba coupling. In $d=2$  the obtained Casimir force displays singular behavior in the limit of vanishing $\nu$.

\end{abstract}

\maketitle

\section{Introduction} 
In recent years the idea of coupling internal degrees of freedom to their momenta has been developed theoretically and experimentally realized in systems of ultracold gases of neutral bosonic atoms \cite{Dalibard_2011, Galitski_2013, Zhou_2013, Goldman_2014, Zhai_2015, Zhang_2016,  Recati_2022}. Such phenomena may be engineered by exposing specific multi-level Bose systems to non-uniform laser fields, which leads to emergence of bosonic quasiparticles of (pseudo)spin 1/2 coupled to their momenta in a diversity of ways. For many-body systems this gives rise to novel and interesting phenomena such as, {\it inter alia}, possible reorganization of the ground state into the so-called supersolid \cite{Boninsegni_579}, or stabilization of the uniform condensate also in dimensionality $d=2$ at temperatures $T>0$ \cite{Jakubczyk_2024_2}. There have been discussions \cite{Stanescu_2008, Barnett_2012, Barnett_2012_erratum, Ozawa_2012_2, Cui_2013, Yu_2013, He_2013, Liao_2014, Roy_2014, Hickey_2014, Mardonov_2015, Liao_2015, Galteland_2016, Wu_2017, Kawasaki_2017, Yang_2021} concerning the admissible forms of the condensate at $T>0$ but, at least for Rashba-type S-O coupling, it seems now established that the uniform (non-supersolid) condensate is the only one possible \cite{Jakubczyk_2024_2}. In this particular case there remain rather few open questions concerning the structure of the bulk phase diagram. We are  however not aware of studies addressing interfacial properties, such as the Casimir effect in systems of this kind. This situation stands in contrast to the standard  Bose gas (with no S-O coupling), for which the Casimir effect was extensively discussed \cite{Martin_2006, Gambassi_2006, Maciolek_2007, Napiorkowski_2013, Jakubczyk_2016, Diehl_2017, Flachi_2017, Faruk_2018, Dantchev_2019, Lebek_2020, Thu_2020, Aydiner_2020, Dantchev_2020, Panochko_2021, Napiorkowski_2021, Bhuiyan_2021, Lebek_2021, Song_2021, Panochko_2022, Pruszczyk_2023}.  
One exception is Ref. \cite{He_2017}, which however focuses on the plane-wave ordered phase, which (for Rashba-type S-O coupling) exists as a thermodynamic phase only at $T=0$. Questions concerning Casimir physics in such setups are interesting for at least three immediate reasons: (i) S-O coupling introduces a new energy scale to the system, which elevates the number of arguments of the scaling functions in question; (ii) the S-O coupling relates to anisotropies which propagate to the macroscopic scales of Casimir phenomena, in other words: in three-dimensional systems the value of the Casimir interaction depends on the relative orientation of its direction and the plane defined by the S-O coupling; (iii) S-O coupling stabilizes 2-dimensional condensates and therefore gives rise to the long-range Casimir effect also in this dimensionality; of particular theoretical interest is the limit of vanishing S-O coupling, where the condensate becomes progressively expelled from the phase diagram, leading to singular behavior of the Casimir energy.   

In this paper we investigate the equilibrium interfacial properties of the basic model of free bosonic particles with Rashba S-O coupling and enclosed in a periodic slit. Focusing on the condensed state, we evaluate and discuss the Casimir energy in dimensionality $d\in \{2,3\}$. We report a behavior richer and different as compared to the standard situation involving structureless, noninteracting  bosonic particles \cite{Martin_2006, Gambassi_2006}. 

The present paper is organized as follows: in Sec.~II we briefly introduce and review the S-O coupled Bose gas and in Sec.~III we discuss its bulk phase diagram. Our new results are presented in Sec.~IV, where we discuss the Casimir effect in $d=2$ (Sec.~IV A) and $d=3$ (Sec.~IV B). The paper is summarized in Sec.~V.  
\section{The S-O coupled ideal Bose gas}
We consider the $d$-dimensional ($d\in \{2,3\}$) gas of non-interacting Rashba S-O coupled bosons described by the Hamiltonian \cite{Stanescu_2008}:
\begin{align}    
\hat{\mathcal{H}}=\sum_{\vec{p},\alpha,\beta}\langle\vec{p},\alpha|\hat{\mathcal{H}}^{(1)}|\vec{p},\beta\rangle \;\hat{a}^\dagger_{\vec{p},\alpha} \hat{a}_{\vec{p},\beta} \;. 
\label{Hamiltonian}
\end{align}
Here $\{\hat{a}^\dagger_{\vec{p},\alpha}\}, \{\hat{a}_{\vec{p},\beta}\}$ denote the bosonic creation/annihilation operators, and  
\begin{equation}
\label{R01}
    \hat{\mathcal{H}}^{(1)}=\frac{\vec{p}^{\,2}}{2m}\hat{\mathbb{I}}_\sigma + v \left(p_1\hat{\sigma}_2-p_2\hat{\sigma}_1\right)\;.
\end{equation} 
The single-particle states are labeled by the momentum $\vec{p}$ and the pseudospin 1/2 index ($\alpha$, $\beta\in\{+,\, -\}$), $\hat{\sigma}_i$ are the Pauli matrices, while $v\geq 0$ denotes the spin-orbit coupling.
 Experimental realizations of setups leading to such artificial pseudospin 1/2 bosons are described, inter alia,  in Refs \cite{Osterloh_2005, Zhu_2006, Satija_2006, Stanescu_2007}.  The system is contained in a hypercubic vessel of volume $V=L^{d-1}\times D$ and we assume periodic boundary conditions in each of the directions. We will be focusing on the scaling regime $\lambda\ll D\ll L$, where $\lambda = h/(2\pi m k_{B}T)^{\frac{1}{2}}$ is the thermal de Broglie length. In $d=3$ distinct situations occur depending on the orientation of the container walls relative to the plane determined by the S-O coupling of Eq.~(\ref{R01}) - See Sec.~IV B.
 
 By applying a unitary transformation $\hat{U}_{\vec{p};\sigma,\beta}$ the Hamiltonian of Eq.~(\ref{Hamiltonian}) can be diagonalized, leading to   
 \begin{align}    
 \hat{\mathcal{H}}=\sum_{\vec{p},\sigma}E_{\vec{p},\sigma} \hat{A}^\dagger_{\vec{p},\sigma} \hat{A}_{\vec{p},\sigma} \,,  
\label{Hamiltonian_d}
\end{align}
 where the operators $\hat{A}_{\vec{p},\sigma} =\sum_\beta\hat{U}_{\vec{p};\sigma,\beta}\, \hat{a}_{\vec{p},\beta}$ fulfil bosonic commutation relations and $\sigma = \pm$. Presence of the S-O coupling ($v>0$) removes the degeneracy with respect to the pseudospin and the corresponding single-particle energy spectrum displays two branches: 
\begin{align}
\label{dispersion}
E_{\vec{p},\pm}=\frac{\vec{p}^2}{2m} \pm v\sqrt{p_1^2+p_2^2}\;. 
\end{align} 
The + branch displays a cusp-like minimum at $\vec{p}=\vec{0}$ with $E_{\vec{0},+}=0$. The - branch exhibits 
a minimum $E_{\vec{p}_0,-}=-mv^2/2$ degenerate on a circle and occurring for $\sqrt{p_1^2+p_2^2}=mv=:p_0$ and $p_3=0$ in the case of $d=3$. 

In the framework of the grand canonical ensemble the free energy $\Omega$ is additive with respect to the two components: 
\begin{equation} 
\Omega(T,\mu_{-},\mu_{+},D,L,x_{0}) = \Omega_{-}(T,\mu_{-},D,L,x_{0}) + \Omega_{+}(T,\mu_{+},D,L,x_{0})\;.  
\end{equation}
Here $\mu_{\pm}$ are the  chemical potentials for quasiparticles corresponding to the two energy branches and we introduced the dimensionless coupling constant $x_0= \left(m\nu^2/2k_{B}T\right)^{1/2}$. For later reference we also define the dimensionless chemical potentials: $s_{\sigma} = \mu_{\sigma}/k_{B}T$. The bulk and surface  free energy densities ($\omega_{b,\sigma}$ and $\omega_{s,\sigma}$, respectively), from which the Casimir force follows \cite{Krech_book, Brankov_book, Kardar_1999, Klimchitskaya_2009, Gambassi_2009, Dantchev_2023}, are obtained in the standard way: 
\beq
\label{omegab}
\omega_{b,\sigma}(T,\mu_{\sigma},x_{0}) = \lim_{D,L \rightarrow \infty} \frac{\Omega_{\sigma}(T,\mu_{\sigma},D,L,x_{0})}{L^{d-1} D} \;,
\eeq 
\beqn
\label{omegas}
&\omega_{s,\sigma}&(T,\mu_{\sigma},D,x_{0}) = \nonumber \\ 
 &=& \lim_{L \rightarrow \infty} \frac{\Omega_{\sigma}(T,\mu_{\sigma},D,L,x_{0}) - D L^{d-1} \omega_{b,\sigma}(T,\mu_{\sigma},x_{0}) }{L^{d-1}} \;.
\eeqn 
The possibility of realizing condensation (or its lack) in the thermodynamic limit follows from analysis of the expressions for bulk particle densities $n_{\sigma}(T,\mu_{\sigma},x_{0}) = - \frac{\partial \omega_{b,\sigma}(T,\,\mu_{\sigma},\,x_0)}{\partial \mu_{\sigma}}$. These densities take different forms depending on $d$ and are below discussed for $d=2$ and $d=3$. 

\section{Bulk properties and condensation}
Using the standard expression for the grand canonical partition function of the noninteracting Bose gas \cite{Kardar_book}
\begin{align} 
    \Xi_0^{(\sigma)}(T,\mu_{\sigma},D,L,x_{0}) = \prod_{\vec{p}}\frac{1}{1-e^{-\beta(E_{\vec{p},\sigma}-\mu_\sigma)}} \;, 
\end{align}
and passing to the thermodynamic limit, we obtain the following formulae for $\omega_{b,\sigma}=-\beta^{-1} \log\Xi_0^{(\sigma)}$: 
\begin{align} 
\label{Xi0minus}
   \beta \omega_{b,-}(T,\mu_-,x_0)=&-\frac{1}{\lambda^d}\left[ g_{\frac{d+2}{2}}(z_-)+ x_0\left(\sqrt{\pi}\,g_{\frac{d+1}{2}}(z_-) + f_1^{(d)}(x_0,z_-)\right) \right] \;,
\end{align} 
and 
\begin{align} 
\label{Xi0plus}
    \beta \omega_{b,+}(T,\mu_+,x_0)= -\frac{1}{\lambda^d}\left[g_{\frac{d+2}{2}}(z_+)- f_2^{(d)}(x_0,z_+) \right]\;,
\end{align} 
where we introduced
\begin{align}
    f_1^{(d)}(x_0,z):=\int\limits_0^{x_0^2} dt \left(x_0^{-1}+t^{-1/2}\right)g_{\frac{d}{2}}(z e^{-t})\;, 
    \label{f1def}
\end{align} 
and 
\begin{align}
    f_2^{(d)}(x_0,z):=\int\limits_0^{\infty}dt\left(1+\frac{t}{x_0^2}\right)^{-1/2}g_{\frac{d}{2}}(z e^{-t})\;. 
    \label{f2def}
\end{align} 
Above we also introduced the fugacities $ z_-:=e^{s_- + x_0^2}\;,\; z_+:=e^{s_+}$\;,
and the Bose functions $g_\alpha(z) :=\sum_{k=1}^{\infty}z^k/k^\alpha$\;. 
Differentiating the above expressions for the free energies yields the densities $n_\pm$: 
 \beqn
\label{Densitieseq1}
n_+(T,\mu_{+},x_{0})\, \lambda^d  =g_{\frac{d}{2}}\left(z_+\right)- f_2^{(d-2)}\left(x_0,z_+\right)+\frac{\lambda^d}{V}\frac{z_+}{1-z_+} , 
\eeqn     
 \beqn
\label{Densitieseq2}
n_-(T,\mu_{-},x_{0})\, \lambda^d =  g_{\frac{d}{2}}\left(z_-\right) + \nonumber \\ \sqrt{\pi}\,x_0\, g_{\frac{d-1}{2}} \left(z_-\right)+x_0f_1^{(d-2)}(x_0,z_-) +\frac{\kappa\lambda^d}{V}\frac{z_-}{1-z_-}\;.  
\eeqn
In thermodynamic states supporting the existence of Bose-Einstein condensates, the last terms on the RHSs of Eqs (\ref{Densitieseq1}) and (\ref{Densitieseq2}) (proportional to $1/V$) correspond, upon properly evaluating the limit $V \rightarrow \infty$ and $z_{\sigma} \rightarrow 1^{-}$,  to the condensates' densities $n_{\sigma,0}$, see \cite{Jakubczyk_2024_2}. The remaining terms yield the thermal densities $n_{\sigma}^{th}(T,\mu_{\sigma},x_{0})$, and $\kappa$ is a numerical constant originating from the degeneracy of the ground state of the $E_-$ band. In complete analogy to the description of condensation in the standard non-interacting Bose gas, in the thermodynamic limit condensation is possible only if the thermal densities are bounded from above (i.e. finite at $z_\sigma=1$). 

  We first observe that, for $x_{0}=0$ both thermal densities reduce to $n_{\sigma}^{th}(T,\mu_{\sigma},0) \lambda^d = g_{d/2}(z_\sigma)$ which diverges for $z_{\sigma} \rightarrow 1^-$ for $d\leq 2$ and remain finite otherwise. This recovers the well known fact regarding the absence of BEC in the standard two-dimensional ideal Bose gas and its presence for $d>2$ for $T$ sufficiently low. 

In case of the density $n_{+}$, Eq.(\ref{Densitieseq1}),  we note that for $x_0>0$ the dominant contribution to the integral defining $f_2^{(d-2)}(x_0,z_+)$ [Eq.~(\ref{f2def})] comes from small values of the integration variable $t$ and the integrand is exponentially suppressed for $t$ large. We expand the integrand in powers of $t/x_0^2$, integrate term by term and observe  the cancellation of the first term of this expansion with  $g_{\frac{d}{2}}(z_{+})$ on the RHS of Eq.~(\ref{Densitieseq1}), which yields:
\beq
\label{n+asympt}
    n_{+}^{th}(T,\mu_{+},x_{0})\, \lambda^d \approx \frac{1}{2x_0^2}\, g_{\frac{d}{2}+1}(z_+) \;.
\eeq
This result leads to the conclusion that the lower critical dimension for condensation of particles from the $E_{\vec{p},+}$ branch is lowered due to the presence of S-O coupling such that it becomes admissible also for $d=2$ (note that the system makes no sense for $d<2$). On the other hand, the expression for $n_{-}^{th}$ in Eq. (\ref{Densitieseq2}) contains the term $g_{\frac{d-1}{2}}(z_-)$ unbounded for $z_-\to 1^-$ as long as $d\leq 3$. This divergence is not cancelled by the term involving $f_1^{(d-2)}$ in Eq. (\ref{Densitieseq2}). In consequence, the condensate of particles from the $E_{\vec{p},-}$ branch (corresponding to the modulated supersolid-type phase) is in fact possible only for $d>3$.  

In summary: in presence of the Rashba S-O coupling, particles from the $E_{\vec{p},+}$ band may condense in dimensionality $d\geq 2$, while the condensate of particles from the $E_{\vec{p},-}$ band (the supersolid phase) is impossible for $d\leq 3$ and $T>0$. We note that the present analysis pertinent to the noninteacting system is almost identical (and leads to equivalent  conclusion) to the one recently performed for an analogous system in presence of interparticle interactions treated at mean-field level \cite{Jakubczyk_2024_2}. We also emphasize that for vanishing $x_0$ the phase diagram of the system is not evolved smoothly to the one corresponding to $x_0=0$ \cite{Jakubczyk_2024, Stachowiak_2025}; for $d=3$ switching on an arbitrarily weak S-O coupling immediately expels one of the condensed phases from the phase diagram, while for $d=2$ an arbitrarily small $x_0>0$ stabilizes the $E_{\vec{p},+}$ condensate.

\section{Casimir effect}
In view of the results of the previous section, the study of the Casimir effect at $T>0$ may be restricted to analysis of the uniform condensate related to the $E_{\vec{p},+}$ band. We note in particular, that oscillations of the Casimir potential reported in Ref.~\cite{He_2017} should occur only at $T=0$, where the plane-wave phase may remain stable. We also point out that for the previously analyzed cases within the BEC phases, the Casimir amplitudes for the perfect and imperfect (mean-field) Bose gases are closely related (see e.g. \cite{Lebek_2021, Napiorkowski_2021}). Our analysis amounts to  calculating the Casimir free energy density  $\omega_{s,+}$ as defined by Eq.~(\ref{omegas}). The Casimir force then follows by taking its (minus) derivative with respect to $D$. We discuss the two cases of $d=2$ and $d=3$ separately. 

\subsection{Case $d=2$} 
For $d=2$, assuming the system's extension in the $x$ and $y$ directions is $L$ and $D$ respectively, and considering periodic boundary conditions, we have: 
\beqn
\label{compfreeenergy}
\frac{\Omega_{+}(T,\mu_{+},D,L,x_{0}) }{k_{B}T} = \nonumber \\
\sum_{n_{x}=-\infty}^{\infty}\sum_{n_{y}=-\infty}^{\infty} \log\left(1-e^{s_{+} + x_{0}^2 - \pi \lambda^2 \left(\sqrt{\frac{n_{x}^2}{L^2} + \frac{n_{y}^2}{D^2}} +  \frac{p_{0}}{h} \right)^2} \right)\;, 
\eeqn  
which in particular, after taking the thermodynamic limit, yields:  
\beq
\label{bulkenergy}
\frac{\omega_{b, +}(T,\mu_{+},x_{0}) \lambda^2}{k_{B}T} = 2 \int\limits_{0}^{\infty} dx \,x \,\log \left(1 - e^{s_{+}-x^2-2 x x_{0}}\right) \;.  
\eeq
Next we insert the above expressions into Eq.~(\ref{omegas}), and perform the limit $L\to\infty$, keeping $D$ fixed. In the subsequent step, we implement the non-expanded form of the Euler-Maclaurin formula 
\beq
\label{summ}
\sum_{n=-\infty}^{\infty} \phi(n) = \int\limits_{-\infty}^{\infty} dx \phi(x) + 2  \sum_{p=1}^{\infty} \int\limits_{-\infty}^{\infty} dx \phi(x) \cos(2\pi p x) 
\eeq
which holds for even functions $\phi(x)$ such that $\phi(\infty) =0$. As a result we obtain the following expression: 
\beqn
\label{omegas2}
\frac{\omega_{s,+}(T,\mu_{+},D,x_{0}) \lambda}{k_{B}T} = \nonumber \\
\frac{4D}{\lambda} \sum_{p=1}^{\infty} \int\limits_{0}^{\infty} dx\, x \log\left(1 - e^{s_{+} - x^2 - 2 x x_{0}}\right) J_{0}(2\pi^{\frac{1}{2}} p \frac{D}{\lambda} x)\;, 
\eeqn
where $J_0$ is the Bessel function. In the BEC phase we have $s_+=0$. After performing the integration variable change $t=2\sqrt{\pi}pxD/\lambda$, assuming $x_0>0$ and $D/\lambda\gg 1$, the argument of the exponential is dominated by the term involving $x_0$ and the other term may be dropped. Expanding the logarithm via $\log(1-x)=-\sum_nx^n/n$, and integrating term by term, leads to the following expression: 
\beq 
\label{omega_s_phi}
\frac{\omega_{s,+} \lambda}{k_{B}T} = -\frac{1}{\pi^{3/2}}x_0\left(\frac{\lambda}{D}\right)^2\varphi\left(\frac{x_0}{\sqrt{\pi}}\frac{\lambda}{D}\right)\;,
\eeq
where 
\beq
\varphi(\tilde{x})=\sum_{p=1}^{\infty}\sum_{n=1}^{\infty}\frac{1}{\left(p^2+n^2\tilde{x}^2\right)^{3/2}}\;.
\eeq  
We observe that for $\tilde{x}\ll 1$ summation over $n$ in $\phi (\tilde{x})$ can be replaced by integration 
and the neglected terms vanish in the limit of $\frac{x_0}{\sqrt{\pi}}\frac{\lambda}{D}$ small. Afer this step both the integral and the remaining summation over $p$ can be carried out analytically. 

We additionally make the observation that $\varphi(\tilde{x}^{-1})=\tilde{x}^3\varphi (\tilde{x})$, which allows us to resolve the limit of $\frac{x_0}{\sqrt{\pi}}\frac{\lambda}{D}\gg 1$ in a completely analogous way.  
This leads to the following asymptotic expressions: 
\begin{align}
\frac{\omega_{s,+} D}{k_{B}T} =\phi(x) \approx \left\{
\begin{array}{lccl}
\label{asymp1}
-\frac{\zeta(2)}{\pi}  + \frac{\zeta(3)}{2\pi^{\frac{3}{2}}} \frac{1}{x} &\equiv \phi_{1}(x) & {\rm for}& x \gg 1  \\ 
-\frac{\zeta(2)}{\pi^{\frac{1}{2}}} x + \frac{\zeta(3)}{2} x^2 & \equiv \phi_{2}(x) & {\rm for} &  x \ll 1\;,
\end{array}
\right.
\end{align} 
where we introduced $x=D/(\lambda x_0)$ and keep $x_{0} \neq 0$; the case $x_{0}=0$ is discussed separetely later on. \\
 Numerical comparison of the full scaling function for the Casimir energy as implied by Eq.~(\ref{omegas2}) and its asymptotic forms 
 $\phi_{1}(x)$ and $\phi_{2}(x)$ valid for $x\gg 1 $ and $x\ll 1$, respectively, Eq. (\ref{asymp1}), is depicted in Fig.~1. 
\begin{figure}
\includegraphics[width = \linewidth]{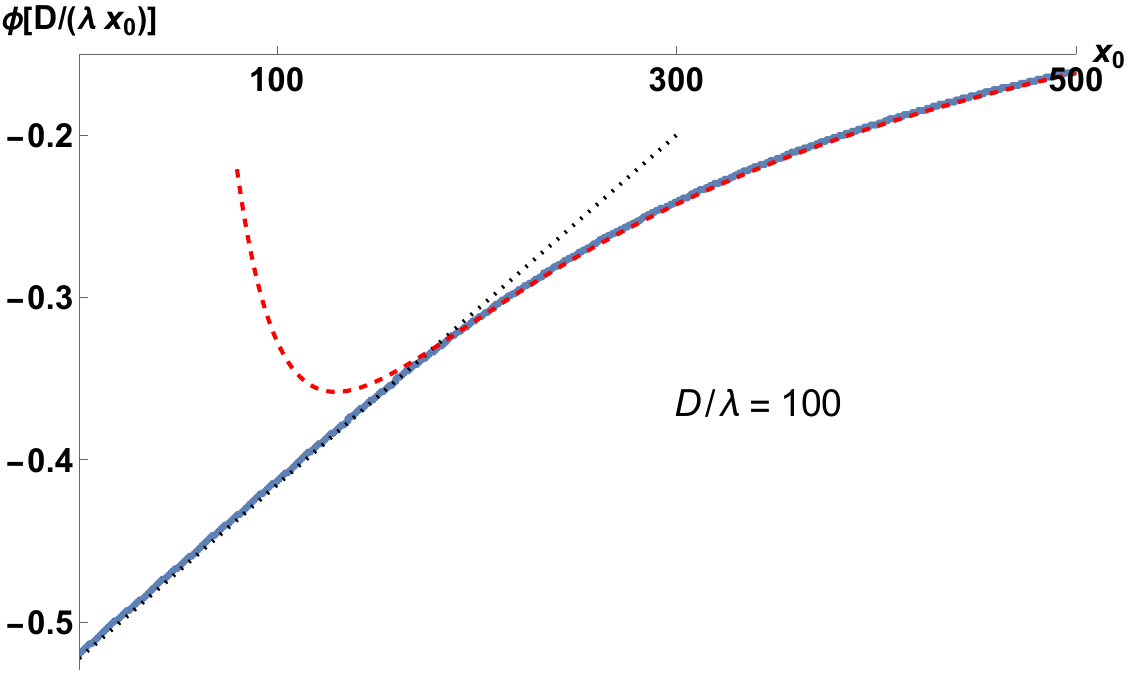}
\caption{The scaling function $\phi=\beta\omega_{s,+}D$ implied by Eq.~(\ref{omegas2}) plotted as a function of $x_0$ at fixed $D/\lambda=100$. For $x_0$ small the function $\phi$ is well approximated by the asymptotic form given by $\phi_1$ (dotted line), while for $x_0$ large, it coincides with the function $\phi_2$ (dashed line). The limit of vanishing $x_0$ is not resolved in this plot scale - compare Fig.~2.  } 
\label{d2_plot}
\end{figure} 
The above scaling behaviour of the surface free energy density implies the following form of the Casimir force 
$F_{Cas, +}(T,\mu_{+},D,x_{0})$ evaluated per unit length at  $\mu_{+}=0$ and for $\frac{D}{\lambda} \gg 1$ 
\begin{align}
\label{casd2}
\frac{F_{Cas, +}(T,0,D,x_{0})}{k_{B}T} = \left\{
\begin{array}{lccl}
\frac{1}{D^2} \left(-\frac{\zeta(2)}{\pi}  + \frac{\zeta(3)}{\pi^{\frac{3}{2}}} \frac{x_{0}\lambda}{D}\right)  & {\rm for} &  & \frac{D}{\lambda x_{0}} \gg 1 \\ 
- \frac{\zeta(3)}{2} \frac{1}{\lambda^2 x_{0}^2}  & {\rm for} &  & \frac{D}{\lambda x_{0}} \ll 1 \;.
\end{array}
\right.
\end{align}
The Casimir force is attractive and decays with increasing $D$. For $\frac{D}{\lambda} \gg 1$ and $\frac{D}{\lambda x_{0}} \gg1$ the decay is $\sim 1/D^2$. For $\frac{D}{\lambda} \gg 1$ and $\frac{D}{\lambda x_{0}} \ll1$ the Casimir force seizes to explicitly depend on the slab thickness $D$ and decays $\sim-1/x_{0}^2$. In this case the Casimir force's decay for increasing $D$ follows from the requirement $\frac{D}{\lambda x_{0}} \ll1$. 

We now discuss the limit $x_0\to 0$, which has to be treated with care, since in our above analytical treatment, we neglected the term $\sim x^2$ in the exponential of Eq.~(\ref{omegas2}). Numerical resolution of the limit of vanishing $x_0$ of Eq.~(\ref{omegas2}) is presented in Fig.~2, where, upon increasing $D/\lambda$, we observe nonuniform convergence of the scaling function to the form envisaged from the asymptotic formula $\phi_1$.    
\begin{figure}
\includegraphics[width = \linewidth]{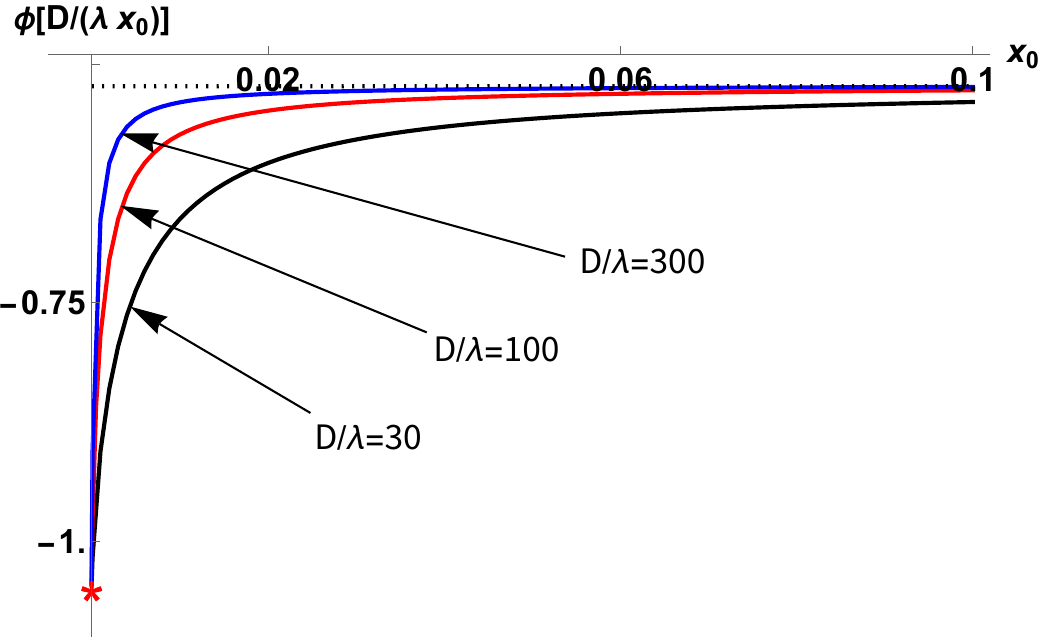}
\caption{The scaling function $\phi=\beta\omega_{s,+}D$ implied by Eq.~(\ref{omegas2}) plotted for a sequence of values of $D/\lambda$ for $x_0$ approaching zero. The dotted line represents the asymptotic form given by $\phi_1$ to which the sequence of plotted functions converges nonuniformly. The red star represents the value $-2\zeta(2)/\pi$ -see the main text.} 
\label{d2_plot_small_x0}
\end{figure} 
The limiting value of the scaling function for $x_0\to 0^+$ can be obtained analytically by putting $x_0=0$ is Eq.~(\ref{omegas2}), which leads to 
\beq
\label{omegas3}
\frac{\omega_{s,+}(T,0,D,0) D}{k_{B}T} = - \frac{2 \zeta(2)}{\pi} \; , 
\eeq  
which fully agrees with the result of numerical evaluation as displayed in Fig.~2.  

Let us summarize the physical picture emerging from the above results. The limits $x_0\to 0$ and $D/\lambda \to\infty$ do not commute. By fixing $x_{0}=0$ and considering sufficiently large $D/\lambda$ one arrives at the limiting value 
$\phi(\infty)= -2\zeta(2)/\pi$. In contrast, by fixing sufficiently large $D/\lambda$ and taking the limit $x_{0} \to 0$ one falls into the regime governed by the function $\phi_{1}(x)$ with $\phi_{1}(\infty)=-\zeta(2)/\pi$.


 Observe however that in the above calculation we always put $\mu_+=0$, which is true in the BEC state, which, at given $T>0$, exists only above a critical density $n_{+,c}(T,x_0) = n_{+}^{th}(T,0,x_{0})$ such that $n_{+,0} = n_{+} - n_{+,c}$. The critical density diverges for $x_0\to 0$ (compare Sec.III). Therefore by considering the limit of small $x_0$, we tacitly assumed that the density $n_{+}$ is adapted such that the BEC state (and in consequence also the long-ranged Casimir interaction) remains supported.  

\subsection{Case $d=3$}
For $d=3$ we distinguish two situations differing in the relative orientation of the S-O coupling plane and the container. The S-O coupling axis is - by definition -  perpendicular to the S-O coupling plane. We refer to the situations, where the system remains finite, i.e. of extension $D$, either in the $z$-direction or in the $x$-direction as "Orientation I" and "Orientation II", respectively. On conceptual level the calculation is analogous to the one performed for $d=2$ and concentrates on evaluating the Casimir free energy $\omega_{s,+}$ as defined by Eq.~(\ref{omegas}), from which the Casimir force follows by differentiation. We discuss the two cases separately.
\subsubsection{Orientation I}
 For the case of S-O coupling axis oriented perpendicular to the confining walls, we obtain:    
\beqn
\label{oms01}
\frac{\omega_{s,\pm}^{(I)}(T,\mu_{\pm},D,x_{0}) \lambda^2}{k_{B}T} = \nonumber \\  
= - \frac{4D}{\lambda} \sum\limits_{p=1}^{\infty} \sum\limits_{n=1}^{\infty} \frac{e^{ns_{\pm}}}{n^{\frac{3}{2}}} \int\limits_{0}^{\infty} dx x e^{-n(x^2 \pm 2 x x_{0}) -\pi \frac{p^2 D^2}{n \lambda^2}} \;.
\eeqn
Since, alike in $d=2$, only the quasiparticles from the $+$ branch condense, we restrict our analysis to this branch and  focus on the BEC state, where $s_+=0$ and the Casimir force is long-ranged. For $x_0=0$  we recover from Eq.~(\ref{oms01}) the known result \cite{Martin_2006}: 
\beq 
\label{Martin}
\frac{\omega_{s,+}^{(I)}(T,0,D,0) \lambda^2}{k_{B}T}= -\frac{\zeta(3)}{\pi}\left(\frac{\lambda}{D} \right)^2\;,
\eeq 
valid for $D\gg\lambda$. 

The situation where $x_0>0$ is entirely different. In full analogy to the previously analyzed case $d=2$ (see the discussion of Sec.III),  one may, for $D\gg\lambda$, disregard the term $x^2$ in the exponential of the integrand in Eq.~(\ref{oms01}). By performing the integration over $x$ , the summation over $n$ (restricting to asymptotically large values of $D$), and finally the summation over $p$, we arrive at the simple expression: 
\beq 
\label{oms051}
\frac{\omega_{s,+}^{(I)}(T,0,D,x_{0}) \lambda^2}{k_{B}T} \approx  
- \frac{3  \zeta(5)}{4\pi^2x_{0}^2} \left(\frac{\lambda}{D}\right)^4\;.
\eeq 
The immediate observation is that the power law governing the decay of the Casimir energy is modified with respect to the standard result of Eq.~(\ref{Martin}), the corresponding force is attractive and decays as $1/D^5$ (thus it vanishes more rapidly as compared to the conventional situation without the S-O coupling where it decays as $1/D^3$). In addition, the S-O coupling enters the Casimir force amplitude via the dimensionless coupling $x_0$, thus restricting its universal character. 
 
 \subsubsection{Orientation II}
The setup, where the S-O coupling axis is oriented parallel to the confining walls is somewhat more complicated. For this situation, by the same sequence of transformations as in the previously discussed cases, the Casimir energy can be cast in the following form: 
\begin{align} 
\label{oms02}
&\frac{\omega_{s,\pm}^{(II)}(T,\mu_{\pm},D,x_{0}) \lambda^2}{k_{B}T} =   \\
&-\frac{4D}{\lambda} \sum\limits_{p=1}^{\infty}  \int\limits_{0}^{\infty} dx x\, g_{\frac{3}{2}}\left(e^{s_{\pm}-x^2\mp 2 x x_{0}}\right)J_{0}\left(2\pi^{\frac{1}{2}}p\frac{D}{\lambda}x\right) \,. \nonumber
\end{align} 
We again restrict our analysis to the $+$ branch and specify $\mu_+=0$. It is easy to show that for $x_0=0$ and $D\gg\lambda$ we recover the standard result of Eq.~(\ref{Martin}). On the other hand, for $x_0>0$ we may again drop the term $\sim x^2$ in the exponential occurring in Eq.~(\ref{oms02}). Subsequently, implementing the series representation of $g_{\frac{3}{2}}$ and integrating term by term, we obtain 
\beqn
\label{oms07}
\frac{\omega_{s,+}^{(II)}(T,0,D,x_{0}) \lambda^2}{k_{B}T} 
\approx    - \frac{D}{\lambda x_{0}^2} \, \sum\limits_{p=1}^{\infty}  \sum\limits_{n=1}^{\infty} \, 
\frac{1}{n^{\frac{1}{2}} \left( n^2 + p^2 \frac{\pi D^2}{\lambda^2 x_{0}^2}\right)^{\frac{3}{2}} } \,.\hspace{1cm}
\eeqn
The asymptotic behavior of $\omega_{s,+}^{(II)}(T,0,D,x_{0})$ for $D/\lambda \gg1$ and $x_{0} \gg 1$ depends on the value of the ratio $D/\lambda x_{0}$. 
For $D/\lambda x_{0} \ll 1$ one may replace the summation over $p$ with an integral, perform the integration and the summation over $n$, which yields:    
\beq 
\label{oms071}
\left.\frac{\omega_{s,+}^{(II)}(T,0,D,x_{0}) \lambda^2}{k_{B}T}\right |_{\frac{D \lambda}{x_{0}} \ll1} \approx  
- \frac{1}{x_{0} \pi^{\frac{1}{2}}} \left( \, \zeta\left(\frac{5}{2}\right) - \frac{\pi^{\frac{1}{2}} D}{2 \lambda x_{0}} \zeta\left(\frac{7}{2}\right) \right)\,. 
\eeq 
Thus, in this regime, the dominant contribution to the Casimir force per unit area does not depend on the distance $D$ and has the form 
\beq
\label{cas072}
\frac{F_{Cas, +}^{(II)}(T,0,D,x_{0})}{k_{B}T} = - \frac{\zeta\left(\frac{7}{2}\right)}{2} \frac{1}{\lambda^3 x_{0}^2}\,.
\eeq 
Note that the similar  (i.e.,  independent of $D$)  behavior of the Casimir force is observed in the analogous regime in $d=2$ (Sec.~IV A).

In order to handle the opposite asymptotic behavior ($D/\lambda x_{0} \gg 1$) by this same technique we first pull out the factor $[\sqrt{\pi}D/(\lambda x_0)]^{-7/2}$ out of the sum in Eq.~(\ref{oms07}), such that each factor of $n$ is  divided by $\sqrt{\pi}D/(\lambda x_0)$. Only then is it legitimate (for $D/\lambda x_{0} \gg 1$) to replace the summation over $n$ with an integral. After performing the resulting integration and the remaining summation over $p$, we obtain the following expression  
\beq 
\label{oms081}
\left.\frac{\omega_{s,+}^{(II)}(T,0,D,x_{0}) \lambda^2}{k_{B}T}\right |_{\frac{D}{ \lambda x_{0}} \gg1} \approx  
- \frac{4\zeta (5/2)\Gamma(5/4)^2}{\pi^{1/2} x_{0}} \,\left(\frac{\lambda x_{0}}{D}\right)^{3/2}\,.
\eeq 
Concentrating on the case $D/\lambda x_{0} > 1$ we additionally evaluated the Casimir energy in question numerically directly from  Eq.~(\ref{oms02}). Our numerical analysis is exemplified in Fig. 3 and fully agrees with Eq.~(\ref{oms081}). 
\begin{figure}
\includegraphics[width = \linewidth]{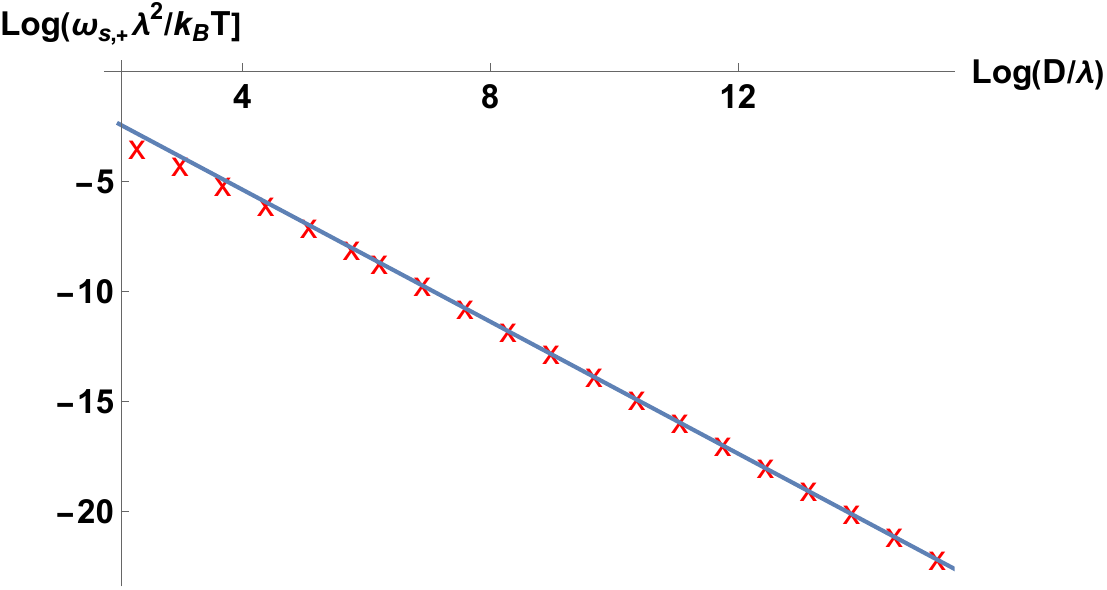}
\caption{Numerical evaluation of the Casimir energy as a function of $D/\lambda$ from Eq.~(\ref{oms02}) (red crosses). The value of $x_0=10$ is fixed. The fit to the data (blue solid line) indicates a power law with the exponent -1.50 - see the main text and Eq.~(\ref{oms081}).   } 
\label{d2_plot}
\end{figure} 
 The resulting Casimir force is attractive and decays as $1/D^{5/2}$. Note that the resulting decay exponent is non-integer in this case and, similarly to the previous case of Orientation I, the S-O coupling enters the Casimir force amplitude via the dimensionless coupling $x_0$. 

\section{Summary}
Presence of the S-O coupling gives rise to significant modifications of the phase diagram and bulk properties of Bose systems. In the present paper we have studied how this effect is reflected in interfacial features represented by the thermal Casimir interactions occurring in presence of Bose-Einstein condensation at $T>0$. The relevance of the S-O coupling is particularly strong in two dimensions, where the Casimir effect is entirely absent  if the S-O coupling is zero.  
We have identified the relevant scaling variable governing the Casimir energy and the related scaling regimes. In all the considered cases, involving different spatial dimensionalities and orientations of the confining walls relative to the S-O coupling axis, the Casimir interaction remains attractive. The corresponding power-law for its decay is however modified. In particular, in $d=3$ and confining walls oriented parallel to the S-O coupling axis the exponent governing this decay is not integer. The degree of universality of the Casimir interaction also becomes restricted, since its magnitude depends on the S-O coupling constant. The present study is concentrated on finite temperatures and the thermodynamic state involving the Bose-Einstein condensate. It might be very interesting to also address the approach towards $T=0$, where the spatially-modulated (supersolid) condensates are expected to be stable. The other relevant open question concerns analysis of types of S-O coupling different than the presently discussed pure Rashba coupling.    

\begin{acknowledgments}
 We acknowledge support from the Polish National Science Center via grant 2021/43/B/ST3/01223. We thank Evgeny Sherman for discussions on related topics and for remarks on the manuscript.
\end{acknowledgments} 
   
\bibliography{bibliography.bib}
\bibliographystyle{apsrev4-1}

\end{document}